# The effect of dust charge variation, due to ion flow and electron depletion, on dust levitation

Victor Land, Angela Douglass, Ke Qiao, Lorin Matthews and Truell Hyde

*The Center for Astrophysics, Space Physics and Engineering Research, Baylor University, Waco, TX, USA, 76798, www.baylor.edu/casper*

**Abstract.** Using a fluid model, the plasma densities, electron temperature and ion Mach number in front of a powered electrode in different plasma discharges is computed. The dust charge is computed using OML theory for Maxwellian electrons and ions distributed according to a shifted-Maxwellian. By assuming force balance between gravity and the electrostatic force, the dust levitation height is obtained. The importance of the dust charge variation is investigated.

## LEVITATION OF DUST AND CHARGE MEASUREMENTS

In experiments at the Center for Astrophysics, Space Physics and Engineering Research (CASPER), complex plasma bi-layers were formed in a modified Gaseous Electronics Conference (GEC) RF discharge in argon, the geometry of which has been described elsewhere [1]. By adjusting the discharge power, the distance between the two layers was affected, as illustrated in figure 1. It is clear that the inter-layer separation is decreased by increasing the discharge power. Figure 2 shows the levitation height above the electrode surface for each particle size, for a similar experiment. In this case, the distance between the layers is decreased to half a millimeter, which is the approximate Debye length for this discharge [2].

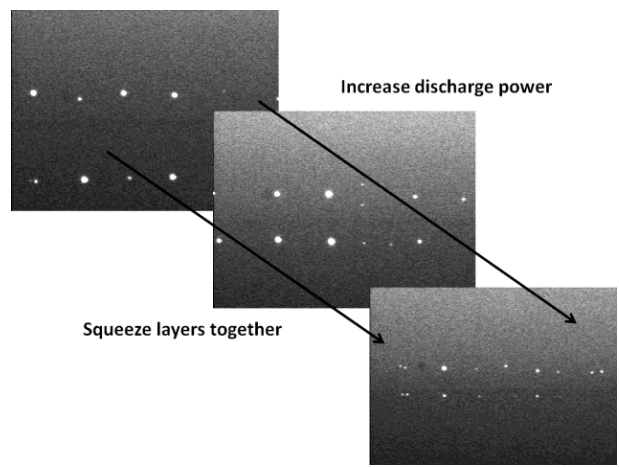

**FIGURE 1.** Three frames illustrating the bi-layer experiments, as discussed in the text. The upper layer consists of 6.5 micron diameter melamine-formaldehyde (MF) particles, while the lower layer consists of 11.9 micron MF dust particles.

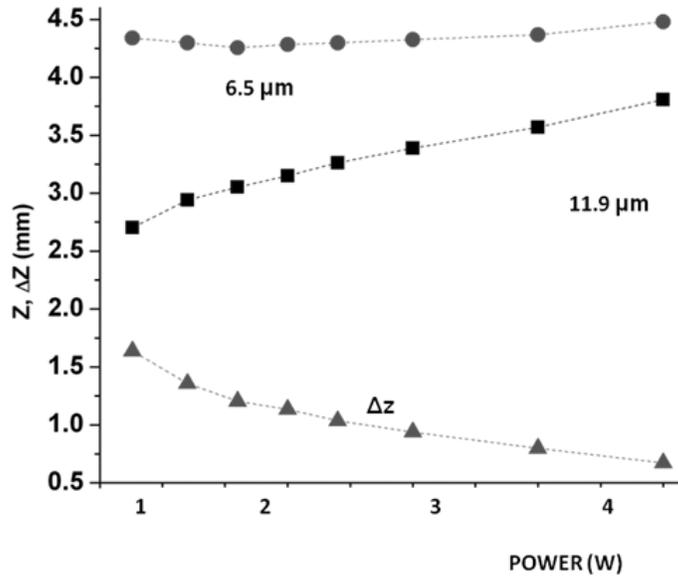

**FIGURE 2.** The levitation height above the powered electrode for the two layers in a complex plasma bi-layer as the applied discharge power is varied. The experiment was performed in argon at 25 Pa.

Even though the intent of the experiment, namely to influence the inter-layer distance, was achieved, upon further investigation, the levitation height of the particles could not be analytically reproduced using Orbital Motion Limited (OML) theory [3] and the usual assumption of a linear electric field [4].

In order to insure the above results were not due to any *bi-layer* interaction, the experiment was repeated for different sizes separately, at a pressure of 20 Pa. These results are shown in figure 3, against the driving potential, $V_{RF}$ (The power is proportional to $V^2_{RF}$). Although the levitation height of the smaller particles does not change much, the larger particles initially move up quickly with increased power, before reaching a constant levitation height.

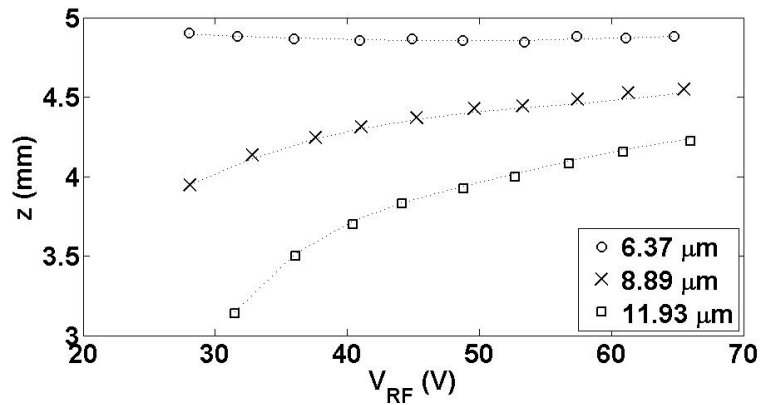

**FIGURE 3.** The levitation height of single layer plasma crystals consisting of MF particles of different diameters, plotted versus the applied amplitude of the driving potential, $V_{RF}$. Each crystal was separately suspended in the discharge, but the results are plotted together.

Recently, measurements of the dust charge in similar discharges were reported using two different methods. The first employed a rotating electrode in which a centripetal force was induced on levitated tracer particles, allowing determination of the dust charge for different particle-sizes [5]. The second was performed by placing the discharge in a gondola attached to a centrifuge, in order to artificially vary the acceleration of gravity perpendicular to the electrode, which moved suspended tracer particles down towards the lower electrode [6].

The first paper reported a more negative charge when compared to OML theory for larger dust particles suspended closer to the electrode surface, indicating an *increase in the dust charge with decreasing levitation height*. It was argued that the increase in ion flow speed towards the electrode surface decreased the positive ion current contribution for the dust charging, leading to the increase in negative charge. The second paper reported an observed *decrease in the dust charge towards the lower electrode*. This apparent contradiction raised our interest, especially given the results presented above. We therefore decided to use a self-consistent fluid model for dusty plasma, for which the complete description can be found elsewhere [7], in order to compute the plasma profiles in our plasma discharges and from these the dust charge and levitation height.

## FLUID MODEL RESULTS AND COMPARISON TO EXPERIMENTS

In order to determine the dust charge on a dust particle levitated in plasma with ion flow, the OML charging currents were calculated for Maxwellian electrons, $I_e$, and positive ions with a shifted-Maxwellian distribution, $I_+$, as given in Eq. 1 and Eq. 2, respectively. At equilibrium, the currents will balance; $I_+ + I_e = 0$, resulting in an equation for the dust charge potential, $\Phi_D$. By solving the profiles for $\alpha = n_+/n_e$, the ion mach number, $M_+ = u_+/V_T$, with $u_+$ the ion drift speed and $V_T = (k_B T_+/m_+)^{0.5}$ the thermal speed, with $T_+$ the ion temperature and $m_+$ the ion mass, and the profile for the electron temperature, $T_e$, the dust potential was obtained for different heights above the powered electrode, for different particle sizes and discharge geometries. Using a capacitor model for the dust particles, the dust charge can now be obtained as $Q_D = -Z_D e = 4\pi\varepsilon_0 a \, \Phi_D$, with $a$ the dust particle radius.

$$I_e = -e\sqrt{8\pi}a^2 \sqrt{\frac{k_B T_e}{m_e}} n_e exp\left(\frac{e\Phi_D}{k_B T_e}\right), \qquad (1)$$

$$I_+ = e\sqrt{2\pi}\sqrt{\frac{k_B T_+}{m_+}} n_+ a^2 \left(\sqrt{\frac{\pi}{2}} \frac{1+M_+^2-2e\Phi_D/k_B T_+}{M_+} erf\left(\frac{M_+}{\sqrt{2}}\right) + exp\left(\frac{-M_+^2}{2}\right)\right). \qquad (2)$$

The results for the depletion profile and ion drift profile in a GEC discharge are shown in figure 4. The electron density decreases by roughly an order of magnitude over the distance from the bulk to the electrode, almost independently of the discharge power (or amplitude of the driving potential). The ion drift is significant and increases linearly with driving potential.

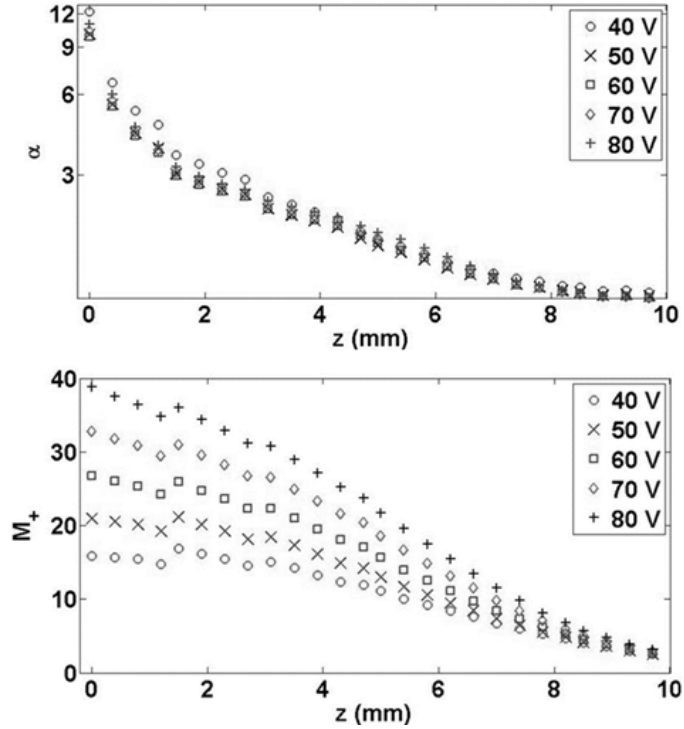

**FIGURE 4.** The ion density over electron density, $\alpha$ (top), showing electron depletion in the sheath, and the ion Mach number, $M_+$ (bottom), showing the acceleration of the ions in the sheath.

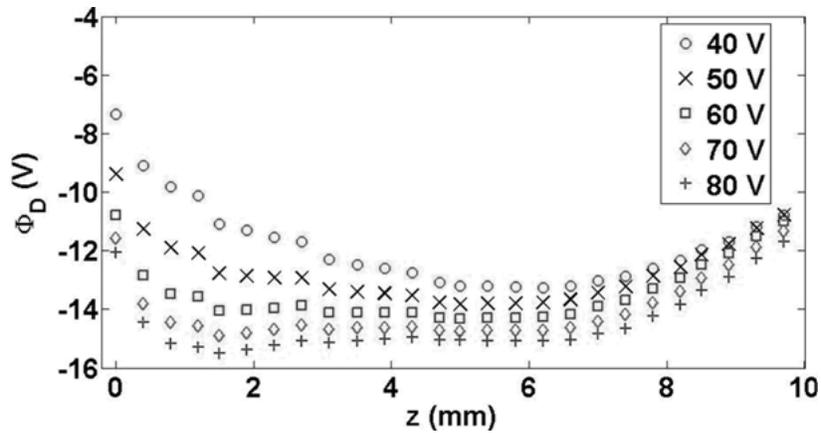

**FIGURE 5.** The dust surface potential profile obtained using the charging equations and profiles obtained with the fluid model, as discussed in the text. It shows a clear maximum at a point between the bulk and the electrode.

Using these profiles and the electron temperature profile (not shown here), the dust floating potential was obtained and is shown in figure 5. The profiles show a clear maximum negative dust charge. During the transition from the bulk to the sheath, the ions are accelerated, but the electron depletion is small, so that the negative dust charge increases. Moving deeper into the sheath, the electron depletion becomes important and the dust charge becomes less negative again.

Employing force balance, the levitation height of the dust particles can be obtained, as shown in figure 6. The qualitative behavior of the dust is the same, namely that the smallest particle sizes remain at the same height with increasing driving potential, while larger particles first increase their height rather quickly, before reaching a constant height. This shows that dust charge variation, due to electron depletion and ion drift, is important for the proper description of the behavior of dust particles levitated in the plasma.

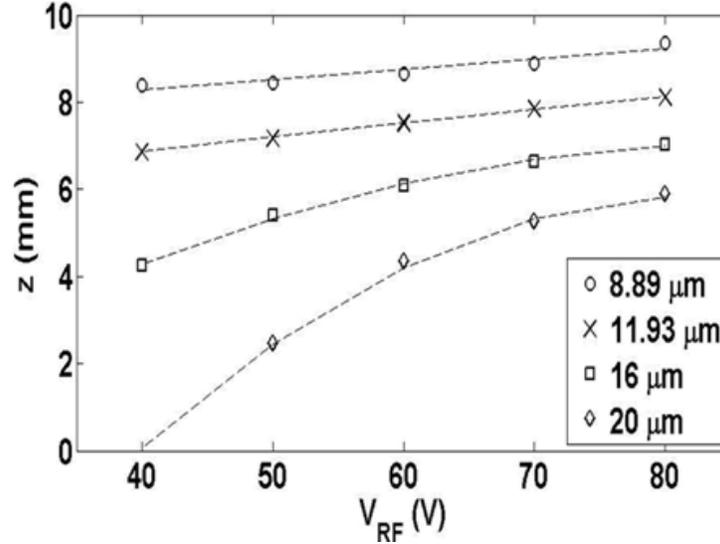

**FIGURE 6.** The dust levitation height obtained with the fluid model, assuming force balance between gravity and the electrostatic force. Qualitatively, similar behavior as in the experiment can be seen.

Applying this model to the geometry used in the rotating electrode method [5], the dust charge was obtained for different sizes, as summarized in table 1. Dust charges for larger particles (deeper in the sheath) can be more negative than expected from OMl theory, as reported in [5], but depending on the discharge parameters, they can also be less negative, when the dust particles are levitated closer to the bulk.

**TABLE 1.** The dust charge ratio for different particle sizes (indicated by the subscript) modeled in a similar geometry as used in the rotating electrode method [5].

| $V_{RF}$ (V) | $Z_{20}/Z_{16}$ | $Z_{20}/Z_{12}$ |
|---|---|---|
| 50 | 1.16 | 1.62 |
| 60 | 1.24 | 1.75 |
| 80 | 1.28 | 1.84 |
| Expected (OML) | 1.25 | 1.68 |

The results of the model when applied to the hyper-gravity experiments [6] are shown in figure 7. Depending on particle size, the dust charge can decrease with gravity, or increase with gravity, depending on the particle position with respect to location of the maximum of the dust charge profile.

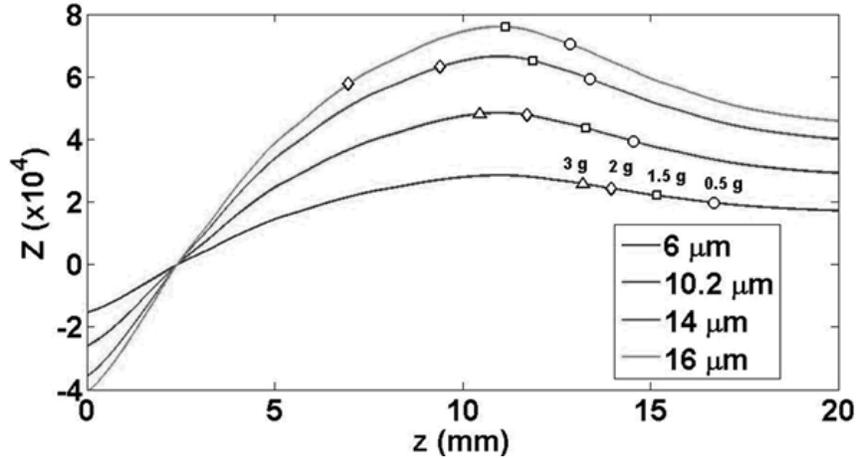

**FIGURE 7.** The dust charge profile for different particle sizes and their levitation height for different effective gravities (indicated by the different open symbols).

## CONCLUSIONS

When taking the ion flow and electron depletion into account in the charging of dust in plasma in front of a powered electrode, a local dust charge maximum is obtained. Depending on dust size and discharge settings, the dust charge can increase or decrease with height. Despite the model's simplicity, for instance lacking any effect of collisionality on the dust charging, it captures most of the qualitative features of different experiments involving the measurement of the dust charge and the levitation height. One should note that this might be important for any region where electron depletion and ion flow is important, for instance for the dust charging near the void edge in complex plasma under microgravity conditions, or quite possibly dust charging in astrophysical dusty plasma systems, where charged particle flow is ubiquitous.

## ACKNOWLEDGMENTS

This work is supported by the National Science Foundation under Grant No. 0847127.